\documentclass{MIPRO}

\usepackage{cite}
\usepackage{amsmath,amssymb,amsfonts}
\usepackage{algorithmic}
\usepackage{graphicx}
\usepackage{textcomp}
\usepackage{xcolor}
\usepackage[T1]{fontenc}
\usepackage{flushend}
\usepackage{multirow}
\usepackage{multicol}
\usepackage{array}
\usepackage{url}
\usepackage{amsmath}
\usepackage{bm}
\usepackage{siunitx}
\usepackage{cleveref}
\usepackage{glossaries}

\newacronym{cn}{CN}{compute node}
\newacronym{hpc}{HPC}{high-performance computing}

\crefname{figure}{Fig.}{Figs.}
\crefname{table}{Table}{Tables}
\crefname{equation}{Eq.}{Eqs.}
\crefname{section}{Sec.}{Secs.}

\graphicspath{{./figs/}}

\newcommand{\CC}{C\nolinebreak\hspace{-.05em}\raisebox{.2ex}{\small\bf +}\nolinebreak\raisebox{.2ex}{\small\bf +}}
\newcommand{\tcpu}{t_\text{CPU}}
\newcommand{\tcomm}{t_\text{c}}

\newcommand{\ccfl}{\Delta t_\text{max}}

\begin{document}

\title{Performance Trade-offs of High Order Meshless Approximation on Distributed Memory Systems}

\author{
\IEEEauthorblockN{
J. Vehovar\IEEEauthorrefmark{1}\IEEEauthorrefmark{2},
M. Rot\IEEEauthorrefmark{1}\IEEEauthorrefmark{2},
G. Kosec\IEEEauthorrefmark{2}
}

\IEEEauthorblockA{\IEEEauthorrefmark{1} 
Jozef Stefan International Postgraduate School, Jamova cesta 39, 1000 Ljubljana, Slovenia}

\IEEEauthorblockA{\IEEEauthorrefmark{2} 
Institut "Jožef Stefan", Parallel and Distributed Systems Laboratory, Jamova cesta 39, 1000 Ljubljana, Slovenia}

jon.vehovar@ijs.si

}

\maketitle

\begin{abstract}
  Meshless methods approximate operators in a specific node as a weighted sum of values in its neighbours. Higher order approximations of derivatives provide more accurate solutions with better convergence characteristics, but they come at the cost of including more neighbours. On the accuracy-per-compute time basis we know that increasing the approximation order is beneficial for a shared memory computer, but there is additional communication overhead when problems become too large and we have to resort to distributed memory systems. Meshless nodes are divided between systems in spatially coherent subdomains with approximations at their edges requiring neighbouring value exchange. Performance optimization is then a balancing act between minimizing the required number of communicated neighbours by lowering the approximation order or increasing it to enable faster convergence. We use the radial basis function-generated finite difference method (RBF-FD) to approximate the derivatives that we use to solve the Poisson equation with an explicit iterative scheme. Inter-system communication is provided by Open MPI, while OpenMP is used for intra-system parallelisation. We perform the analysis on a homogenous CPU-based cluster where we examine the behaviour and attempt to determine the optimal parameterisation with the goal of minimizing the computational time to reach a desired accuracy.

\end{abstract}

\renewcommand\IEEEkeywordsname{Keywords}

\section{Introduction}

Real world applications, such as hydrodynamical modeling~\cite{Schaye2023}, often require highly detatailed simulations to accurately model systems of interest. These can demand more processing power or memory resources than a single shared memory computer can provide, necessitating the distribution of computations across multiple systems \cite{Verkaik2021}.
To distribute workloads across multiple computers, new synchronization methods are required, and existing algorithms designed for shared memory systems must be adapted to function in parallel due to the distributed nature of the \gls{hpc} systems.

Many physical systems can be described by partial differential equations, but few are simple enough to be solved exactly with closed form solutions.  Consequently, numerous numerical approaches have been developed to approximate these solutions. In such methods, space must be discretized either by points or volume elements to facilitate computation. One of the possible numerical approaches are meshless methods whose major advantage is the ability to be able to compute solutions on scattered nodes, whereas other methods may require meshing or regular distributions of points with predefined relations.

Meshless methods approximate function derivatives at a point by appropriately weighing function values in its immediate neighbourhood. Reliance on a limited number of close points makes them well-suited for distributed memory systems. A part of the point set that discretizes the space can be allocated to each system and then systems exchange values that are needed by a particular system, but are not unique to it \cite{Angulo2014}.

Higher order approximations provide better rate of convergence\footnote{With increasing discretization accuracy, i.e. decreasing distance between nodes, a higher order approximation will approach the exact value faster.} but require more neighbouring nodes to evaluate. For the radial basis function-generated finite differences (RBF-FD) method~\cite{tolstykh2003using} it has already been shown by Jančič~\textit{et al.}~\cite{Jancic2021} that the increased rate of convergence of the higher order approximation outweighs the additional computational costs when high accuracy is required. However, it remains unclear whether the same also holds for distributed memory systems where latency and bandwidth limitations impose additional communication cost. The aim of this paper is to investigate how the additional complexity impacts the trade-offs of increasing the approximation order and how the distributed algorithm scales.

The derivative approximation method, domain decomposition, implementation details, and test problem of choice are presented in \ref{sec:methods}. The benchmarking approach, results, and analysis are discussed in \ref{sec:results}, where we analyse the error dependence and efficiency of the algorithm.

\section{Methods}
\label{sec:methods}

Finite difference methods \cite{Sirca2025} compute function derivatives on a structured grid of nodes by summing appropriately weighted contributions of values at locations of neighbouring nodes. Meshless methods offer the same capability, but generalised to the unstructured positioning of nodes.

In general, they approximate a linear operator $\mathcal{L}$ acting on a function $u$ at a node $\bm{x_c}$,
\begin{equation}\label{eq:mm_derivative_eval}
 \mathcal{L}u(\bm{x}_c) = \sum_{i = 1}^{n} w_i u(\bm{x}_i),
\end{equation}
where $w_i$ are weights and $\bm{x}_i$ are nodes in the neighbourhood of $\bm{x}_c$~\cite{leborne2023guidelinesRBFFD}. These nodes are said to be in the support of $\bm{x}_c$ and $n$ is referred to as the support size. Usually they are chosen as the $n$ nearest nodes to $\bm{x}_c$ with $\bm{x}_c$ itself being a member of its support.

The weights can be computed using the RBF-FD~\cite{tolstykh2003using} method where we obtain them by solving the system
\begin{equation}
 \begin{pmatrix}
 \text{A} & \text{P} \\
 \text{P}^\text{T} & 0
 \end{pmatrix}
 =
 \begin{pmatrix}
 \bm{w}\\
 \bm{\lambda}
 \end{pmatrix}
 =
 \begin{pmatrix}
 \mathcal{L}\bm{\Phi}\\
 \mathcal{L}\bm{p}
 \end{pmatrix},
\end{equation}
where
\begin{equation}
 \tiny{
 	\text{A} =
 	\begin{pmatrix}
 	\Phi_1\left(\bm{x}_1\right) \dots \Phi_1\left(\bm{x}_n\right)\\\
 	\vdots\\
 	\Phi_n\left(\bm{x}_1\right) \dots \Phi_n\left(\bm{x}_n\right)
 	\end{pmatrix},
 \quad \bm{w} =
 \begin{pmatrix}
 w_1\\
 \vdots\\
 w_n \end{pmatrix},
 \quad \mathcal{L}\bm{\Phi} =
 \begin{pmatrix}
 \mathcal{L}\Phi_1(\bm{x}_c)\\
 \vdots\\
 \mathcal{L}\Phi_n(\bm{x}_c)
 \end{pmatrix}
}
\end{equation}
The elements of A are defined as $\Phi_i(\bm{x}) = \phi\left(||\bm{x} - \bm{x}_i||\right)$,
where $\phi$ are radial basis functions, with cubic polyharmonic splines ($\phi(r) = r^3$) being a common choice. The monomial augmentation $\text{P}^\text{T}$ and $\mathcal{L}\bm{p}$ is analogous to A and $\mathcal{L}\bm{\Phi}$, respectively, where $\Phi_i$ is replaced by monomials up to the order $m$. For a given $m$ there are $M = \binom{m + d}{d}$ monomials, where $d$ is the number of spatial dimensions, making P a $n \times M$ matrix. 

The order of the derivative approximation is determined by the order of the monomial augmentation $m$~\cite{leborne2023guidelinesRBFFD} with higher orders yielding more accurate derivatives at the cost of greater computational complexity. A popular choice for the support size is $n = 2M + 1$~\cite{bayona2017}, corresponding to $n \in \{13, 31, 57\}$ for $m \in \{2, 4, 6\}$ in 2D.

The accuracy of the solution can be increased either by increasing the order of the linear operator approximation or by increasing the node density. Both increase the computational cost, and the greater the desired solution accuracy, the greater $m$ is generally required to minimise the time to solution, as shown in \cite{Jancic2021}.

\subsection{Domain Decomposition}

A natural choice for parallelization of our problem is to assign nodes (as described at the beginning of \cref{sec:methods}) that sample the original domain to subdomains and assign each of these subsets to a \gls{cn} which performs the iteration of each step in parallel. The data to be exchanged between the \glspl{cn} are the values of $u$ at nodes that are in the support of nodes belonging to a particular subdomain but not contained within it. Therefore, each \gls{cn} is assigned a subdomain set which contains nodes that are unique to the subdomain and those that will require values of $u$ to be exchanged.

Time of data exchange can be, to the first order, approximated with
\begin{equation}\label{eq:comm_time}
 t_\text{e} = \lambda + \frac{D}{B},
\end{equation}
where $\lambda$ is the inherent latency of the network, $D$ is the amount of data sent and $B$ is the bandwidth. Since $\lambda$ and $B$ are properties of the underlying hardware and thus out of our control, the only way for us to affect the communication time is by reducing the amount of data communicated between the \glspl{cn}. Communication minimisation can be achieved by dividing the domain into subdomains where the nodes are contiguous in the sense that they are in each other's support. This leads to a collection of subdomain sets that exchange values with neighbouring subdomain sets at their edge nodes.

\begin{figure}
  \centering
  \includegraphics[width=\columnwidth]{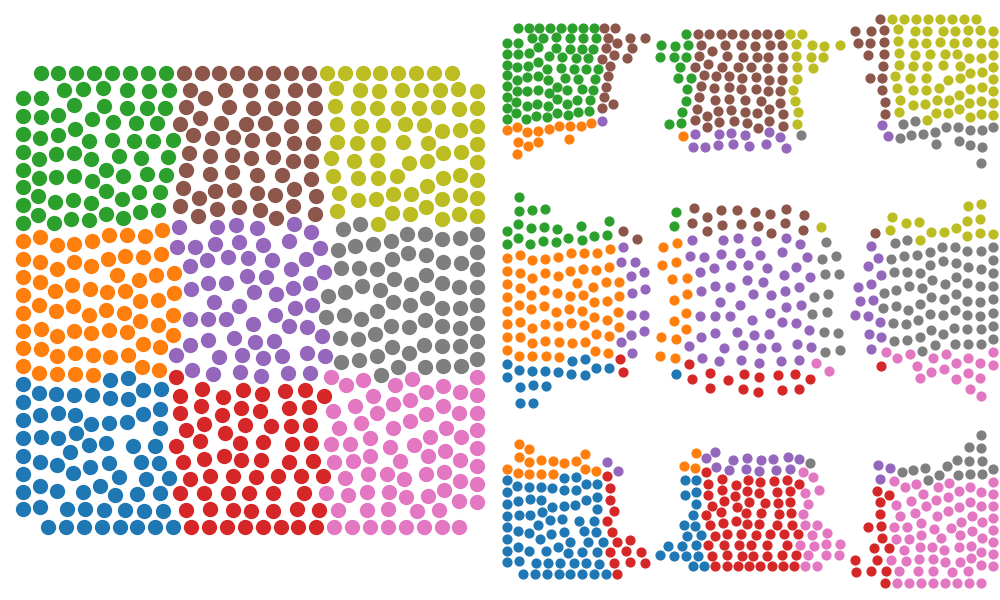}
  \caption{\label{fig:subdomain_demo} An example of the decomposition of the original domain discretization into nine subdomains.}
\end{figure}

We decompose the domain by subdividing it a certain number of times at regular intervals in each dimension. The nodes are then assigned to a subdomain according to which domain subdivision they occupy. An example of a subdivision for a domain with \num{602} nodes and a second order augmentation is shown in \cref{fig:subdomain_demo}.

\subsection{Implementation}

The part of the algorithm concerning local solution of the PDE is implemented in the \CC{} open source Medusa library \cite{Slak2021}. The addition of the domain decomposition is implemented using Open MPI \cite{MPIF2023}, while the parallelisation within the \gls{cn} is enabled by OpenMP \cite{openmp45} with \texttt{\#pragma omp parallel for}.

Our distributed memory system is a homogeneous cluster of 18 \glspl{cn} connected with gigabit ethernet to a single switch, forming a star topology. Each \gls{cn} has two sockets populated with Intel Xeon E5520 CPUs. These have 4 physical cores with 2 threads per core and a single NUMA node per CPU. Each \gls{cn} has \SI{12}{\gibi\byte} of memory available.

The program runs with $N_\text{MPI} = N + 1$ MPI processes (hereinafter referred to as process(es)) (preferably with one process per \gls{cn}), where $N$ is the number of subdomains. The additional process is referred to as the root and does not contribute to the main iteration, but is only used to collect relevant data from the $N$ child processes.

The role of the root process at initialisation is to determine how values are exchanged between the subdomain sets. This is done by iterating over each node in the domain and characterising the nodes in its support relative to the subdomain to which the original node belongs. These nodes can then either be unique to the subdomain or are to be communicated by a process to which a neighbouring subdomain set belongs. An example of such a decomposition can be found in \cref{fig:subdomain_demo}, where colours indicate the affiliation of different nodes to subdomains. A map is required to determine at which nodes values must be communicated between one subdomain set to another.

After this information has been passed on to the child processes, the root's role changes to monitoring. The child processes exchange data asynchronously after each iteration, while the data is sent using MPI's data type \texttt{MPI\_Datatype}, which specifies the values at which the indices of the array $u$ of the subdomain node set are to be sent and received.

\subsection{Test Problem}
\label{sec:test_problem}
We observe the performance of our algorithm on a simple Poisson problem with mixed boundary conditions. Our domain $\Omega$ is a $d$-dimensional cube with a side length of unity where we solve the following problem
\begin{align}
 \Delta u &= f \quad \text{in} \quad \Omega \quad \text{and}\\
 u &= 0 \quad \text{on} \quad \partial\Omega
\end{align}
\begin{equation}
 \text{where} \quad u(\bm{x}) = \prod_{i = 1}^d\sin\left(\pi x_i\right) \quad \text{and} \quad f = - \pi^2du.
\end{equation}

This problem has Dirichelt boundary conditions, but we can also solve the same problem with mixed boundary conditions by solving the problem on only $1/2^d$ of the original volume on a cubic domain with a side length of $1/2$ and choosing the derivative at the boundary that is in the interior of the original domain as $\bm{g} = \bm{0}$. For example, for $d = 2$ we only solve the problem in one quadrant and set the Neumann boundary condition at the sides that lie inside the original square.

There are many superior methods that can be used to solve the given problem, but the Jacobi method~\cite{Press2007} was chosen for its simplicity, as its role is simply to gauge the accuracy of the underlying spatial derivative approximation. Edge node communication times are  proportional for any method using RBF-FD approximation of the same order of accuracy, making these results generalizable. The method iteratively improves the solution at each node in the interior of the domain from step $n$ to step $n + 1$ as
\begin{equation}\label{eq:jacobi_step}
 u(\bm{x}_i)^{n+1} = u(\bm{x}_i)^n + \Delta t \left(\Delta u(\bm{x}_i)^n - f(\bm{x}_i)\right)
\end{equation}
where $\Delta t$ is limited by the stability condition \cite{Press2007}
\begin{equation}\label{eq:cfl}
 \Delta t < h^2 / (2d) = \ccfl
\end{equation}
where $h$ is the node spacing. This condition is derived for regularly distributed nodes and can only serve as a guideline for our irregular distribution. Empirically, we found that $\Delta t = 0.3\ccfl$ is reliably stable, but still of the order of the maximum stable step, which depends on the randomly generated discretisation.

Neumann boundary conditions are enforced at the boundary nodes before starting iteration \eqref{eq:jacobi_step}. The values of $u$ at relevant boundary nodes are computed so that the condition is satisfied. There is no single way to explicitly enforce this condition as satisfying the condition exactly for all relevant boundary nodes at the same time requires the solution of a global system over $\partial \Omega$. For stability reasons and in an effort to keep the domain decomposition simple, we choose to compute the boundary values $u$ by considering only the values from the last step at interior nodes in the support of a given boundary node. This method requires updated values at the boundary nodes that are not unique to a subdomain, but are in the support of nodes within the subdomain. Therefore, the original idea of exchanging the required values at nodes from neighbouring subdomains has to be modified to also communicate values at nodes that are in the support of these exterior boundary nodes\footnote{This can be seen in \cref{fig:subdomain_demo} as a perturbation of the exchanged nodes at a subdomain node set's border.}.

To determine the quality of the solution, we compute its error with respect to the analytical solution $u^a$
\begin{equation}\label{eq:error}
 e = \frac{1}{N_\Omega}\sum_{i = 1}^{N_\Omega} \left|u_i - u^a_i\right|,
\end{equation}
where $N_\Omega$ is the number of nodes that discretise $\Omega$.

We sample the domain by generating a distribution of points of uniform density using a modified version of Poisson disc sampling described in \cite{Slak2019}. This method yields a quasi-uniform distribution of points by starting from a set of seed points $\bm{p_i$} (those that sample the boundary of the domain). The algorithm iterates over the points and generates candidate points around each one at a distance $h(\bm{p}_i)$. The point $\bm{p}_i$ is then removed from the candidate set, while the candidates are added iteratively if they do not violate the proximity conditions and lie within the domain.

\section{Results}
\label{sec:results}

\subsection{Timing setup}
\label{sec:benchmark}

Each benchmark run of the solution to the Poisson problem described in \cref{sec:test_problem} measures the execution time of certain parts of the code. We measure the computation and communication time separately for each process and each step and store the average values for the entire run for each process separately. Since our algorithm does not perform load balancing, care must be taken to synchronise all processes before measuring the communication time, otherwise systematic deviations would occur.

Processes communicate asynchronously via the MPI request handles. First, they start communication via their request handles and then they use them to wait. The wait call blocks until all communication has been completed, including the send calls, which block until the sent data has been received, preventing any overwriting of the sent data in case the sending operation isn't buffered.

At regular intervals, all processes transmit data to the root such as the unique part of their solution and the residual. This is not included in the computation and communication timing, so it has no influence on the results.

\subsection{First Results}
\label{sec:first_results}

\begin{figure}
  \centering
  \includegraphics[width=\columnwidth]{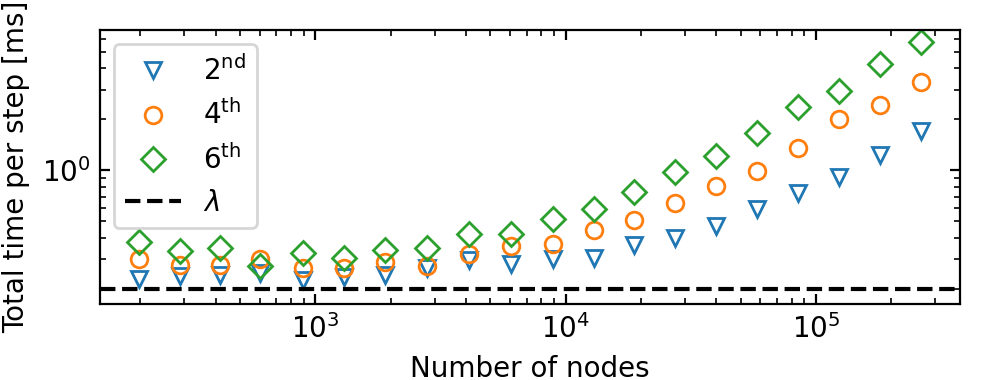}
  \caption{Time per step with respect to the number of nodes in the domain for a configuration of 16 \glspl{cn} and a few different orders of monomial augmentation. The horizontal dashed line shows the order of magnitude of latency.}
  \label{fig:initial_results}
\end{figure}

The overall performance of the algorithm is shown in \cref{fig:initial_results}, which shows the time per step as a function of the number of nodes discretising the domain for a 16-process configuration. The figure compares how the step time varies depending on the choice of the number of nodes discretising the domain and the choice of the order of augmentation. For domains with \num{e4} or more nodes, the scaling regime is nominal, i.e. increasing the number of nodes corresponds to an increase in the total step time. At lower numbers of nodes, the step times start to approach the order of magnitude of the latency $\lambda$. It was estimated by fitting \cref{eq:comm_time} to benchmark data of the communication time of individual processes and the amount of data they send to other processes, and was determined to be about \SI{0.21}{\milli\second}. This is only a rough estimate as the communication pattern is likely to be more complex. Firstly, we would need to account for the additional latency introduced by our synchronisation method, as described in \cref{sec:benchmark}. Secondly, the number of receiver processes varies for each process, so the coordination of communication might introduce additional dynamics that make it difficult to fully account for the delay. Nonetheless, the horizontal line at least serves to indicate the order of communication delay, which could be interpreted as latency.

\subsection{Strong Scaling}

\begin{figure}
  \centering
  \includegraphics[width=\columnwidth]{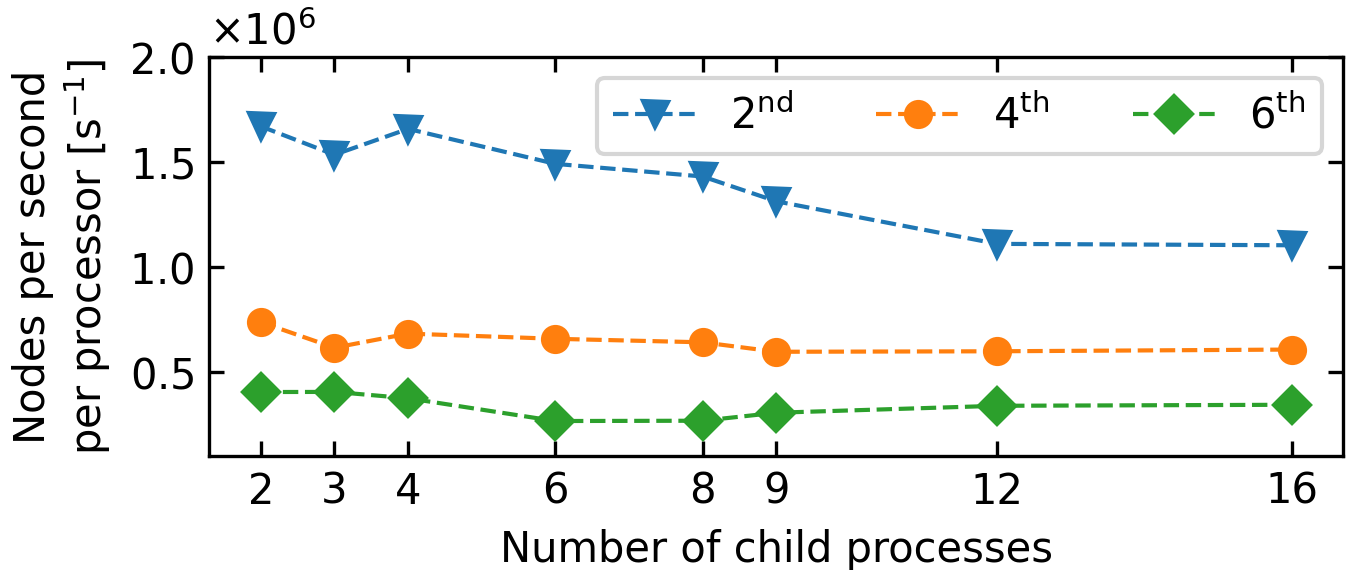}
  \caption{\label{fig:mipro_efficiency_base} Strong scaling in the form of node throughput per CPU as a function of the number of MPI processes for different orders of augmentation.}
\end{figure}

First, we analyse how the strong scaling (fixed problem size of about \num{260000} nodes) is influenced by the choice of augmentation order. The results of this analysis are shown in \cref{fig:mipro_efficiency_base} where the plot shows how the number of nodes processed per second per CPU (hereinafter referred to as node throughput) changes with different choices of augmentation order and the number of processes. The quantity of node throughput is analogous to the efficiency without normalisation. To account for variability, each point is the average of five runs.

The computation time $\tcpu$ for a constant support size is proportional to the number of nodes in the subdomain as
\begin{equation}
 \tcpu \propto \frac{N_\Omega}{p},
\end{equation}
where $p$ is the number of processes. In a regime where latency is not dominant, the communication time is determined by the number of nodes that a process must communicate. If we again assume that these nodes form a thin shell around the unique nodes of the subdomain, the communication time $\tcomm$ is proportional to the subdomain's surface $S$ as
\begin{equation}
 \tcomm \propto S \propto L^{d - 1} \propto V^{1 - 1/d} \propto \left(\frac{N_\Omega}{p}\right)^{1 - 1/d},
\end{equation}
where $L$ is the order of the scale of the subdomain and $V~\propto~L^d$. Therefore, the ratio between the two times follows
\begin{equation}\label{eq:comm_strong_scaling}
 \frac{\tcomm}{\tcpu} = \left(\frac{N_\Omega}{p}\right)^{-1/d}
\end{equation}

As expected, node throughput decreases with increasing augmentation order, and in general the same is true when the number of processes increases due to \eqref{eq:comm_strong_scaling}. The only peculiarity is a slight increase in node throughput when moving from 12 to 16 child processes, which is common to all orders. One possible explanation for this phenomenon are cache effects, but without further investigation this is difficult to confirm.

\subsection{Time and Error Optimization}

\begin{figure}
  \centering
  \includegraphics[width=\columnwidth]{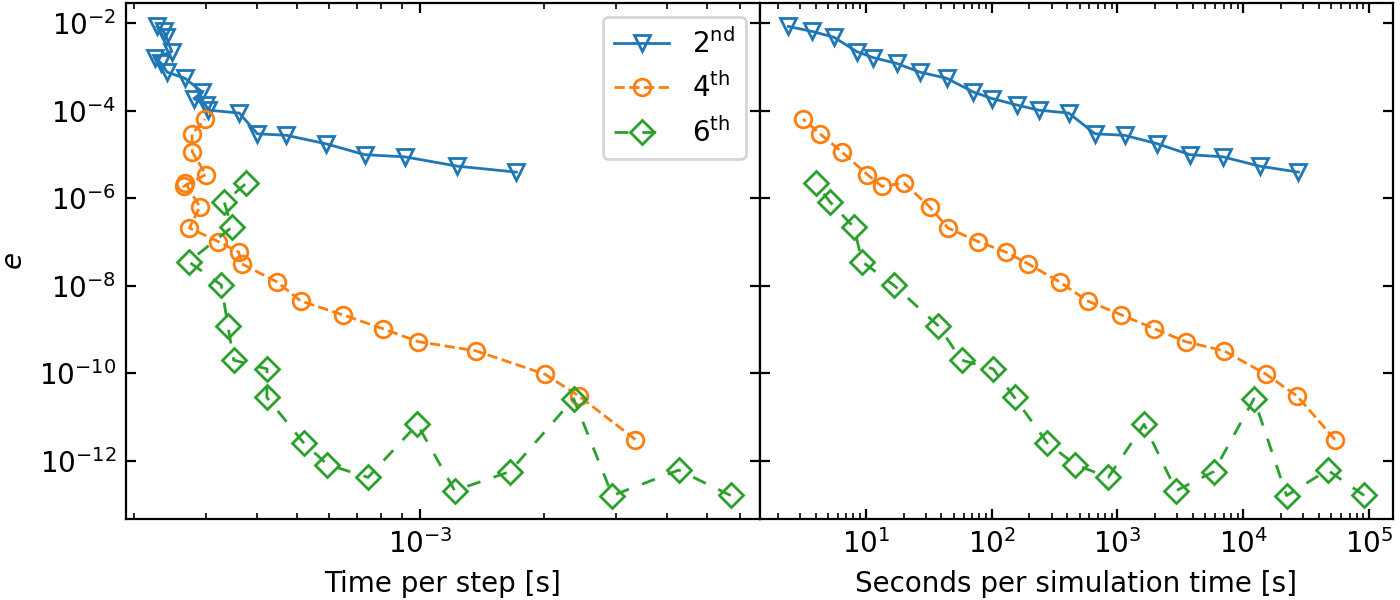}
  \caption{\label{fig:mipro_step_time_vs_err} Parameter scan of monomial augmentation order and node spacing for a configuration with 16 processes.}
\end{figure}

The main interest of this paper is to find the optimal parameters, such as augmentation order and the number of nodes in a domain, to achieve a desired solution accuracy. To do this, we scanned the parameter space of augmentation orders, domain decomposition configurations, and internodal distances. A portion of the results from this scan is shown in \cref{fig:mipro_step_time_vs_err}, where the right panel displays the error as a function of time per step. It is evident that increasing the augmentation order significantly improves the accuracy of the final solution while only modestly increasing step time.

\begin{figure}
  \centering
  \includegraphics[width=\columnwidth]{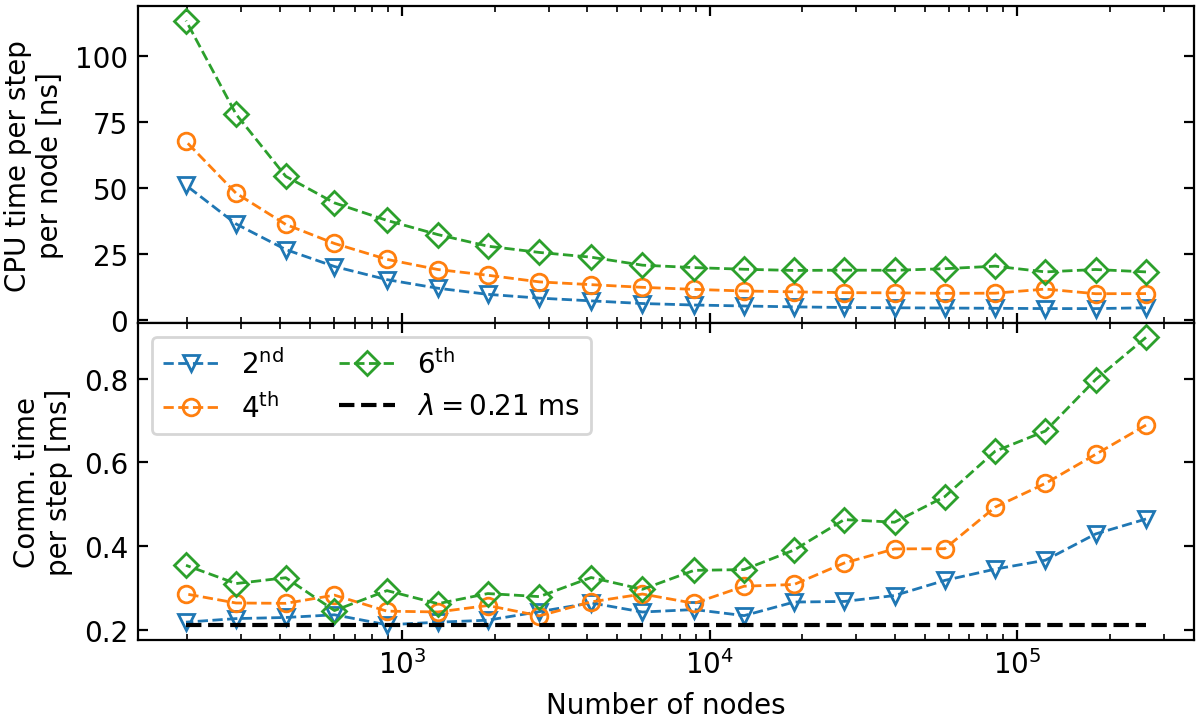}
  \caption{\label{fig:mipro_speedup_2panel} Compute time per step per node and communication time per step for a few different orders of monomial augmentation for a decomposition with 16 processes.}
\end{figure}

Crossover occurs for the lowest times per step, indicating that there might be an advantage to using lower-order augmentations for small problem sizes. However, the usefulness of the distributed algorithm at this problem size is severely limited as can be seen from \cref{fig:mipro_speedup_2panel} where the bottom panel shows the communication time per step for a domain decomposition with 16 processes and a few different augmentation orders. The horizontal dashed black line indicates the order of latency. For problem sizes of \num{e4} nodes or less, this becomes the dominant component of communication time.

In addition to approaching the latency-limited regime, the top panel of \cref{fig:mipro_speedup_2panel} indicates that we are also approaching the regime where the overhead of OpenMP shared memory parallelisation begins to play a role. This is shown by the increase of the computation time required to process one node per step. When the number of nodes in the domain is high, this time is roughly constant as the computational work dominates the time cost, but at some point there are so few computational operations to perform that the overhead associated with work distribution becomes significant.

While these results suggest that increasing the augmentation order when more accurate results are required is preferred to increasing the number of nodes, it can't clearly be shown as the number of steps required to reach convergence is dependent on the chosen internodal distance that determines the maximum stable step size as described by \eqref{eq:cfl}. Regardless of the chosen $\Delta t$, if it is below the stability limit, the progression in simulation time is similar for any reasonable choice of internodal distance. If we assume that convergence will be reached at simulation time $T$, we can obtain the ratio between $T$ and the wall clock time $t_T$ as
\begin{equation}
 t_T = t_\text{ws} \frac{T}{\Delta t} = t_\text{ws} \frac{T}{\alpha \ccfl} \quad \text{or} \quad \frac{t_T}{T} = \frac{t_\text{ws}}{\alpha \ccfl},
\end{equation}
where $t_\text{ws} = \tcpu + \tcomm$ is the wall clock step time and $\alpha = 0.3$, as chosen in \cref{sec:test_problem}.

The right panel of \cref{fig:mipro_step_time_vs_err} shows this derived quantity in place of $t_\text{ws}$ and gives a much clearer performance comparison. It seems like there exist problem sizes where using lower augmentation orders would be advantageous, but such problem sizes would be absurdly small. There also seems that an intersection would be reached between the fourth and sixth order augmentation at low error, but this is just due to the limits of numerical precision. In \cite{Jancic2021} there is some overlap in performance for different augmentation orders for small problems, but this might be more difficult to achieve on distributed systems due to network latency and the positive effects of increasing the support size discussed later in \cref{sec:support_efficiency}. In this case, latency represents a significant amount of time when no processing is taking place, so the advantage of lower order augmentations on small problems which take little time to execute is overshadowed by the time it takes for the network to respond.

\subsection{Efficiency and Support Size}
\label{sec:support_efficiency}

\begin{figure}
  \centering
  \includegraphics[width=\columnwidth]{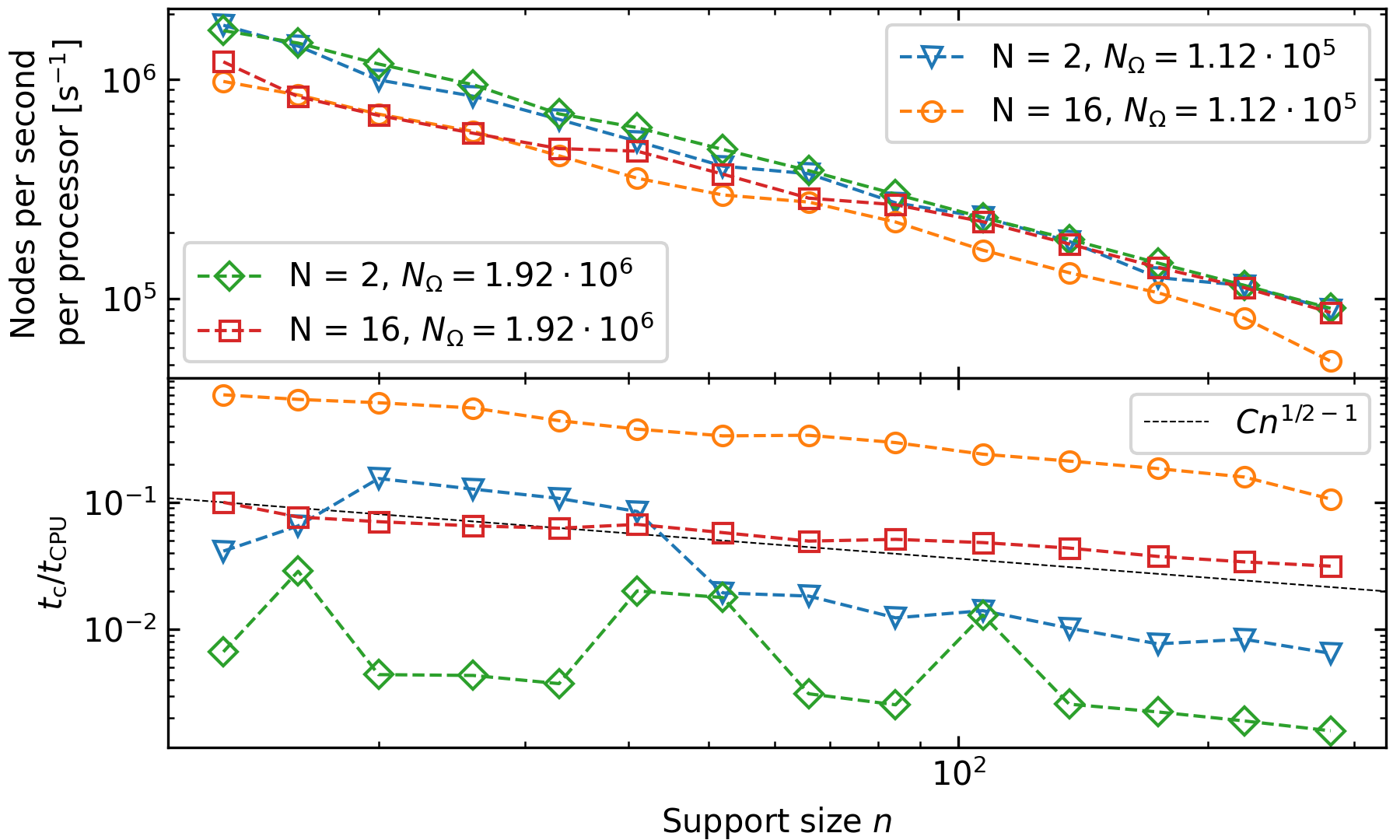}
  \caption{\label{fig:mipro_efficiency} Node processing speed per CPU core for a second order augmentation with increased support size and the corresponding ratio between compute and communication time.}
\end{figure}

The size of the support can be determined by factors other than the augmentation order. Increasing the support size can also help with stabilisation~\cite{Tominec2025}, therefore it is worth investigating how increasing the support size beyond $2M + 1$ influences the performance. A parameter sweep is shown in the top panel of \cref{fig:mipro_efficiency} which shows how the number of processed nodes per CPU core changes as the number of support nodes increases for two different domain decompositions and problem sizes. The curves for two processes are very similar and have a higher throughput for a given support size than those for 16 processes. This indicates that these configurations have converged to an efficient regime. On the other hand, the curve for 16 processes and the smaller problem size consistently performs worse than its two-process counterpart. However, the performance of the 16-process configuration for the larger problem improves, counter-intuitively, as the support size increases and joins the curves for 2 processes at high $n$.

This phenomenon can be explained if we consider how compute and communication times scale with the size of the support. For a constant problem size, the former scales as
\begin{equation}\label{eq:cpu_support_scaling}
 \tcpu \propto n,
\end{equation}
since each derivative evaluation is a dot product, as shown in \cref{eq:mm_derivative_eval}. On the other hand, the communication cost, neglecting latency, is proportional to the number of nodes at the edge of the subdomain that are communicated. If we assume that these nodes form a thin shell, the amount of data sent $D$ is proportional to the shell's thickness which is, which in turn is determined by the radius within which the support nodes are contained. The volume that this radius determines is linearly proportional to the number of support nodes as
\begin{equation}
  n \propto V \propto r^d \quad \Rightarrow \quad \tcomm \approx \frac{D}{B} \propto r \propto n^{1/d}.
\end{equation}
Finally, we have
\begin{equation}
 \frac{\tcomm}{\tcpu} = n^{1/d - 1},
\end{equation}
meaning that for all cases but $d = 1$ increasing the support size decreases the fraction of communication time, thereby increasing the efficiency.

An experimental verification of this scaling is shown in the bottom panel of \cref{fig:mipro_efficiency} where a black dashed line shows the predicted slope. Qualitatively, the results for the 16-process configurations seem to match the predictions quite well, but a comparison for the two-process configurations is more difficult. For each problem size, one could draw two lines parallel with the predicted slope which would have the majority of points (the only exception is the second smallest support size for the smaller problem) near one of them. The offset between these two lines might a be due to the communication specifics between two processes being observable, while this is hidden by the implicit averaging for the 16-process cases.

\section{Conclusions}
\label{sec:conclusions}

We have investigated the effects of choosing different orders of monomial augmentation and support size for explicit solving of partial differential equations using the meshless RBF-FD method on a distributed memory system.

We found that, from accuracy versus computational time perspective, increasing the augmentation order is beneficial. Compared to shared memory computers, where there is some overlap in performance between different approximation orders for small problem sizes, the latency present in communication further suppresses this advantage. This conclusion is further supported by analysis of compute and communication time scaling with support size where the latter scales slower than the former, ensuring that efficiency increases when more accurate solutions are required, as long as the radius of the support remains below the order of the size of the subdomain. Therefore we conclude that on distributed memory computers it is beneficial to use higher order methods, regardless of the problem size, as long as the error introduced by finite difference arithmetic~\cite{Jancic2021} isn't significant.

In future work we would like to confirm the performance benefits of using high-order methods on real world examples and further investigate the odd behaviour in strong scaling for high $N_\text{MPI}$ seen in \cref{fig:mipro_efficiency_base} and $\tcomm / \tcpu$ for $N_\text{MPI} = 2$ seen in \cref{fig:mipro_efficiency}.

\section*{acknowledgment}

The authors would like to acknowledge the financial support of the
Slovenian Research Agency (ARRS) research core funding No. P2-0095 and
the Young Researcher programmes PR-10468 and PR-13389.

\bibliographystyle{IEEEtran}
\bibliography{references}

\end{document}